\documentclass[12pt,english]{iopart}
\usepackage[T1]{fontenc}
\usepackage[latin9]{inputenc}
\usepackage{graphicx}
\usepackage{esint}
\usepackage[dvips]{epsfig}

\makeatletter
\usepackage{iopams}
\usepackage{setstack}

\usepackage{iopams}
\usepackage{amsthm}
\usepackage[dvips]{epsfig}

\makeatother

\usepackage{babel}

\begin{document}

\title{SWNT-array resonant gate MOS transistor}

\author{A. Arun$^{1}$, S. Campidelli$^{2}$, A. Filoramo$^{2}$, V. Derycke$^{2}$,
P. Salet$^{1}$, A. M. Ionescu$^{1}$ and M. F. Goffman$^{2}$}

\address{$^{1}$NanoLab, Ecole Polytechnique Fédérale de Lausanne, CH-1015,
Lausanne, Switzerland}

\address{$^{2}$Laboratoire d'Electronique Moléculaire, SPEC (CNRS URA 2454),
IRAMIS, CEA, Gif-sur-Yvette, France}

\ead{marcelo.goffman@cea.fr}
\begin{abstract}
We show that thin horizontal arrays of single wall carbon nanotubes
(SWNTs) suspended above the channel of silicon MOSFETs can be used
as vibrating gate electrodes. This new class of nano-electromechanical
system (NEMS) combines the unique mechanical and electronic properties
of SWNTs with an integrated silicon-based motion detection. Its electrical
response exhibits a clear signature of the mechanical resonance of
SWNTs arrays (120-150 MHz) showing that these thin horizontal arrays
behave as a cohesive, rigid and elastic body membrane with a Young
modulus in the order of 1-10 GPa and ultra-low mass. The resonant
frequency can be tuned by the gate voltage and its dependence is well
understood within the continuum mechanics framework. 
\end{abstract}


\maketitle
It is largely admitted that the future integration of nano-objects
within technologically relevant devices and circuits will require
their co-integration with the existing CMOS technology. Yet, examples
of functional devices and circuits combining the two types of building
blocks are extremely scarce. Among the different nano-objects, carbon
nanotubes (CNTs) have attracted much attention due to their unique
electronic properties. Examples of integration of CNTs with CMOS-based
electronic circuits are limited to field effect transistors (FET)
\cite{Tseng2004} and interconnects \cite{Close2008}. Moreover, since
CNTs are light and have very large Young's modulus \cite{Treacy96}
they constitute an ideal building block for devices such as micro
and nano-electromechanical systems (MEMS/NEMS). Most of the efforts
to date have primarily focused on individual CNT based nanodevices.
For example, a suspended nanotube has been used as a device element
of a tuneable electromechanical oscillator \cite{Sazonova2004,Van der zant2006},
sensors \cite{Kong2000,Stampfer2006}, rotational actuators \cite{fennimore2003}.
Similarly, a carbon nanotube cantilever has been used in scanning
probe microscope tips \cite{Dai1996}, nanotweezers \cite{nanotweezers},
switches \cite{dujardin2006,switches} and relays \cite{relay} and
even a nanotube radio \cite{nanoradio}. These demonstrations mainly
require dedicated nanofabrication steps for aligning the measuring
circuit to the position of individual CNTs and/or tedious and time
consuming manipulation of CNTs, making the realization of reliable
and integrated nanodevices impractical. Recently an elegant approach
was demonstrated to circumvent this problem, by synthesizing three-dimensional
CNT medium composed mostly of highly-oriented and closely packed CNTs
\cite{Hayamizu2008}. Using this approach, thick {}``CNT wafers''
can be deposited on top of a prefabricated silicon wafer and post-processed,
paving the way to the integration of hybrid CNT-silicon devices. In
order to assess the potential of {}``CNT wafers'' as micromechanical
material, Hayamizu \emph{et al.} \cite{Hayamizu2009} investigated
the mechanical properties of thick beams and found that they act as
a single cohesive unit described by classical theory of elasticity.
On the opposite limit, thin SWNTs arrays used in SWNT-based field
effect transistors showed very promising high frequency electronic
properties \cite{Louarn2007}. If in addition these extremely thin
SWNT layers show interestring mechanical properties they could serve
as a basic material for innovative MEMS/NEMS devices. Of particular
interest are movable gate \cite{Abele2005,Durand2008} and body FET
transistor structures \cite{grogg2008,grogg2008b} operating as active
MEM/NEM resonators, similarly to the very first proposed resonant-Gate
transistor \cite{Nathanson1967}.

Here we present an alternative way, versatile and well adapted to
integrate extremely thin SWNTs arrays to CMOS transistors and demonstrate
the first resonant SWNTs array suspended gate FET, fully compatible
with any bulk silicon CMOS technology. The SWNTs suspended gate FET
device (SWNTs SG-SiFET) combines the virtues of SWNT suspended arrays
(stiff and light material used as vibrating gate of a silicon field
effect transistor (SiFET)), with the integrated transistor detection.
The high frequency vibration of the SWNTs array modulates the charge
density in the FET channel and the output signal of the resonator
is the drain current. The electrical response shows the signature
of the mechanical resonance of SWNTs arrays (120-150 MHz) demonstrating
that these extremely thin horizontal arrays behave as a cohesive,
rigid and elastic body membrane. The resonant frequency can be tuned
by the gate voltage and its dependence is well understood within the
continuum mechanics framework.

\begin{figure}[!htbp]
\centering
\includegraphics[width=13cm]{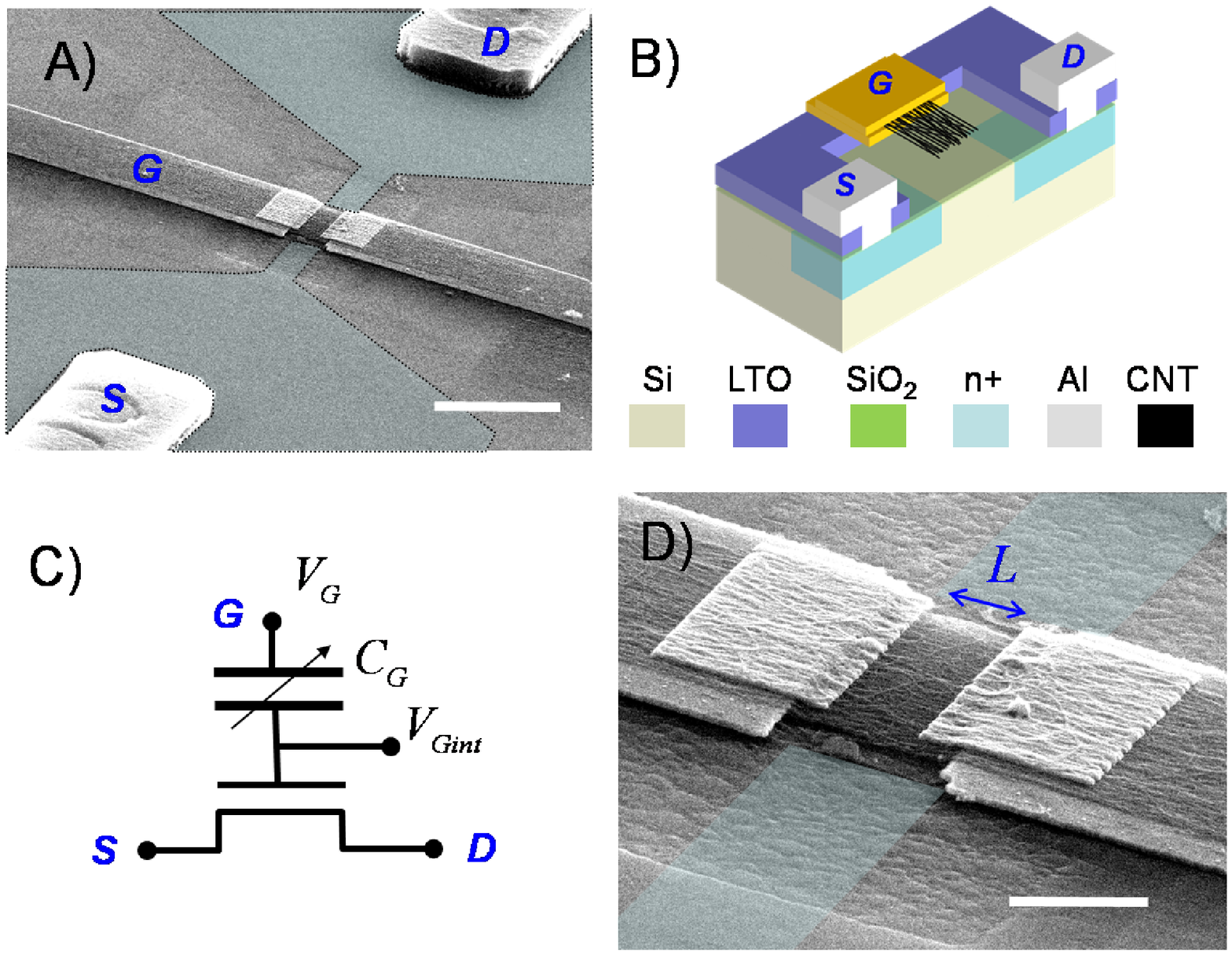}
\caption{\label{fig1}A) Micrograph of a SWNTs SG-SiFET. Scale bar: $5\mu$m.
Light blue: ion implanted regions. B) Schematic cross section of a
SWNTs SG-SiFET device: Si: silicon wafer (p-type; resistivity: 0.1
- 0.5 Ohm cm). LTO: low temperature silicon dioxide deposited to reduce
the parasitic capacitance component from the substrate. n+: ion implanted
regions to form source(S) and drain(D) contacts. Al: S and G contacted
with aluminium Al lines are 50-Ohm designed. CNT: SWNT array deposited
by DEP (see details in the text). C) Electrical circuit of a SWNTs
SG-FET. $C_{G}$ is the SWNTs gate capacitance per unit length. The
SWNTs gate forms a thin SWNTs membrane on top of the FET channel which
can be actuated by applying a voltage difference between the gate
and the channel . D) Detailed view of the SWNTs array. Scale bar:
$1\mu$m.}

\end{figure}

Figure\ref{fig1} shows a micrograph of a SWNTs SG-SiFET and its schematic
cross section. The fabrication details of the SiFET are described
elsewhere \cite{Arun2009}. Suspended SWNTs gates centered on the
FET channel were fabricated as follows: First, chromium/platinum Cr/Pt
(5 nm/ 60 nm) lines were patterned perpendicular to the FET channel
(See \ref{fig1} A and B). These electrodes were used to deposit a
dense array of SWNTs, mostly centered on top of the SiFET channel,
by a dielectrophoresis (DEP) process. The DEP step is described in
detail elsewhere \cite{Louarn2007}. The SWNTs used were synthesized
by laser ablation, first purified and then dispersed at low concentration
in \emph{N}-methylpyrrolidone using moderate sonication, resulting
in a highly stable dispersion comprising mostly individual nanotubes.
To give a precise geometry to the SWNTs gate, an e-beam lithography
step followed by reactive ion etching ($\mbox{O}_{2}$/S$\mbox{F}_{6}$)
was realized. Finally Ti/Au (10 nm$\,$/$\,$50 nm) electrodes were
patterned by e-beam lithography and lift-off technique to doubly clamp
and electrically contact the SWNTs gate to metallic lines. The SWNTs
gate was suspended by etching the sacrificial layer in BHF. This step
was done only in the central region, through an e-beam patterned PMMA
mask, to avoid damaging of Al electrodes of the SiFET. The e-beam
mask was stripped in acetone followed by critical point drying. The
final air-gap of the fabricated device is $~$100nm, with a residual
oxide thickness of $~$35 nm.

The SWNTs array gate is actuated by applying a $V_{G}$ voltage on
the suspended SWNTs membrane gate. The electrostatic force on the
SWNTs membrane for $V_{G}=V_{G}^{DC}+v_{G}\cos\omega t$ and $V_{G}^{DC}\gg v_{G}$
can be well approximated by \begin{equation}
F_{el}\simeq\frac{1}{2}C_{G}^{'}V_{G}^{DC*}\left[V_{G}^{DC*}+2v_{G}\cos\omega t\right]\label{eq:Fel}\end{equation}

where $C_{G}^{'}$ is the first derivative of the gate capacitance
per unit length with respect to the gap separation and $V_{G}^{DC*}=V_{G}^{DC}-V_{Gint}$
and $V_{Gint}$ the mean value of the channel potential \cite{CommentVGint}.
The first term of \eref{eq:Fel} corresponds to the DC component
that elastically deforms the SWNTs membrane and sets its mechanical
tension and the second drives its motion at frequency $\omega$. The
SWNTs membrane motion induced by the AC component of $F_{el}$ modulates
the capacitance $\delta C_{G}$ which in turn modulates the charge
in the channel. The conductance channel change $\delta G$ can be
written as \begin{equation}
\delta G\simeq\frac{dG}{dV_{G}}\left(v_{G}\cos\left(\omega t\right)+\frac{\delta C_{G}}{C_{G}}\left(\omega\right).V_{G}^{DC*}\right)\end{equation}

\begin{figure}[!htbp]
\centering
\includegraphics[width=13cm]{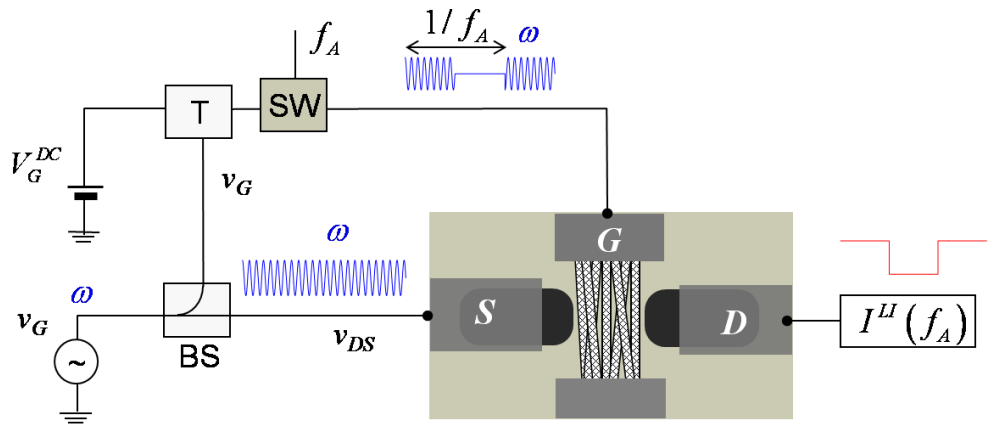} 
\caption{\label{fig2}Diagram of the measurement setup. BS: beam splitter.
SW: GaAs TTL switch. The device is operated as a mixer. A single rf
generator provides two AC voltage signals: one amplitude modulated
by SW (operated at $f_{A}$ ) and applied to the gate electrode to
drive the suspended nanotube array and a second applied to the S electrode
($v_{DS}$). A DC gate voltage is added via a bias-T (indicated by
the {}``T\textquotedblright{}). Its role is twofold: modify the operating
point of the FET and set the mechanical tension of the SWNTs array.
At the drain electrode D, the mixing current has a spectral component
at $f_{A}$ which is measured with a lock-in amplifier. }

\end{figure}

To detect this conductance change at $\omega$ we use the device as
a mixer: we apply an AC voltage $v_{DS}.\cos\left(\omega t-\varphi\right)$
between source and drain \cite{Comment} and an AC gate voltage $v{}_{G}.\cos\left(\omega t\right)$
chopped by a switch operated at audio frequency $f_{A}$ (typically
133 Hz, see setup depicted in Figure \ref{fig2} ).

The source-drain current has a component $I^{LI}$ at $f_{A}$ which
is detected by a lock-in amplifier:

\begin{equation}
I^{LI}=\frac{\sqrt{2}}{\pi}\frac{dG}{dV_{G}}\left(\frac{\delta C_{G}}{C_{G}}V_{G}^{DC*}+v_{G}\right)\cdot v_{DS}\end{equation}

The first term contains the mechanical response of the SWNTs array.
For small motional amplitudes, the membrane can be treated as a simple
harmonic resonator \cite{Yurke1995} with effective mass $M=\xi\rho WLH$
, $\xi\cong1.44858$ and $\rho$, $W$, $L$ and $H$ are the mass
density, width, length and thickness of the membrane respectively.
The change in capacitance $\delta C_{G}\left(\omega\right)$ can be
well approximated by $\delta C_{G}\left(\omega\right)\cong\tilde{C}_{G}^{'}\cdot\delta y\left(\omega\right)$
where $\delta y\left(\omega\right)$ is the displacement of the midpoint
of the resonator and can be obtained from the effective harmonic oscillator
equation (see \Eref{eq:harmonic} below):

\begin{equation}
\delta y\left(\omega\right)=\frac{\tilde{C}_{G}^{'}V_{G}^{DC*}v_{G}}{M}\left[\mathrm{Re}Y\left(\omega\right)\cos\omega t-\mathrm{Im}Y\left(\omega\right)\sin\omega t\right]\end{equation}

Where $Y=\left(\omega_{0}^{2}-\omega^{2}+j\omega_{0}\omega/Q\right)^{-1}$
is the response function of a harmonic oscillator with resonant frequency
$\omega_{0}$ and quality factor $Q$, driven at frequency $\omega$.
Then the lock-in current reads:\begin{equation}
I^{LI}=I_{B}\left(\, A\left(\mathrm{Re}Y\cdot\cos\varphi+\mathrm{Im}Y\cdot\sin\varphi\right)+\cos\varphi\right)\label{eq:I-LI}\end{equation}

Where $I_{B}=\sqrt{2}/\pi\left(dG/dV_{G}\right)v_{DS}v_{G}$ , $A=\frac{\left(\tilde{C}_{G}^{'}\right)^{2}}{M\tilde{C}_{G}}\left(V_{G}^{DC}-V_{Gint}\right)^{2}$and
the phase difference between the RF signal on the gate and source
electrodes is $\varphi$. Using this expression, data can be fitted
and the gate voltage dependence of the resonant frequency $f_{0}\left(V_{G}^{DC}\right)=\omega_{0}/2\pi$
and the quality factor $Q$ can be estimated.

\begin{figure}[!htbp]
\centering
\includegraphics[width=15cm]{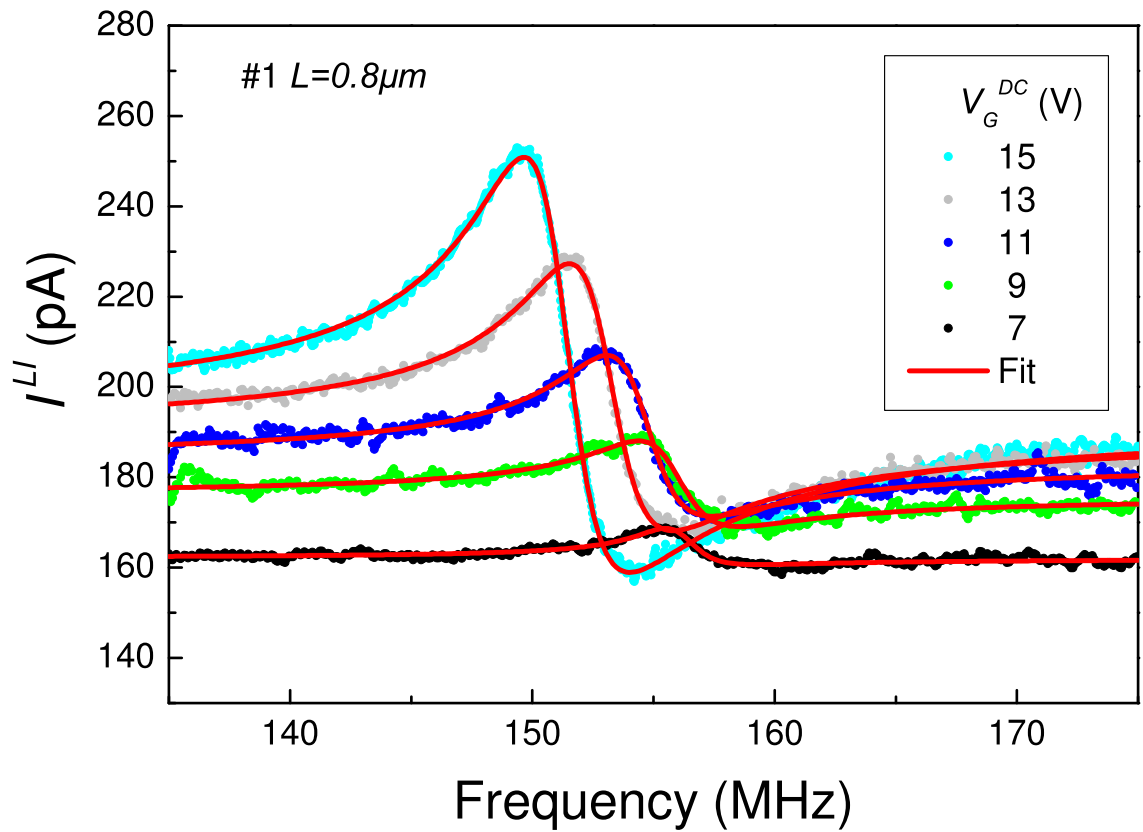}

\caption{\label{fig3}Mixing current $I^{LI}$ as a function of the driving
frequency for different values of $V_{G}^{DC}$ for device \#1. AC
amplitudes are: $v_{G}=-29\mbox{dBm}$ and $v_{DS}=-24\mbox{dBm}.$
Red curves: Best fit using \Eref{eq:I-LI}.}

\end{figure}

Figure \ref{fig3} shows the mixing current $I^{LI}$ measured on
device \#1 ( $L=800\: nm$) as a function of the excitation frequency
for 5 different values of $V_{G}^{DC}$ and their best fit using expression
\eref{eq:I-LI} and $I_{B}$, $A$, $\omega_{0}$, $Q$ and $\varphi$
as fitting parameters. Data were taken in the linear regime (low AC
amplitudes) under vacuum ($10^{-5}$ mbar) and at room temperature.
The size of the resonance feature increases with increasing the DC
gate voltage making evident the increase of the SWNTs membrane deflection
amplitude. The position in frequency of the resonance (in the 150-160
MHz range) moves downward with the increase of the gate voltage showing
an counter-intuitive effective softening of the SWNTs membrane spring
constant. Thanks to the array configuration, the output signal level
of our SWNTs array resonator is approximately one order of magnitude
higher than the one reported in Ref.{[}4{]} and {[}28{]}, using a
mixing configuration scheme, which highlights the advantages of a
hybrid SWNTs/SiFET configuration for building MEM oscillators. Moreover,
the output signal level of the SWNTs array SG-SiFET can be further
increased by operating the transistor at higher DC $V_{DS}$ (saturation
mode). In this work, we mainly focus on the electromechanical behavior
of the SWNTs array and in demonstrating the basic operating behavior
of such device.

\begin{figure}[!htbp]
\centering
\includegraphics[width=13cm]{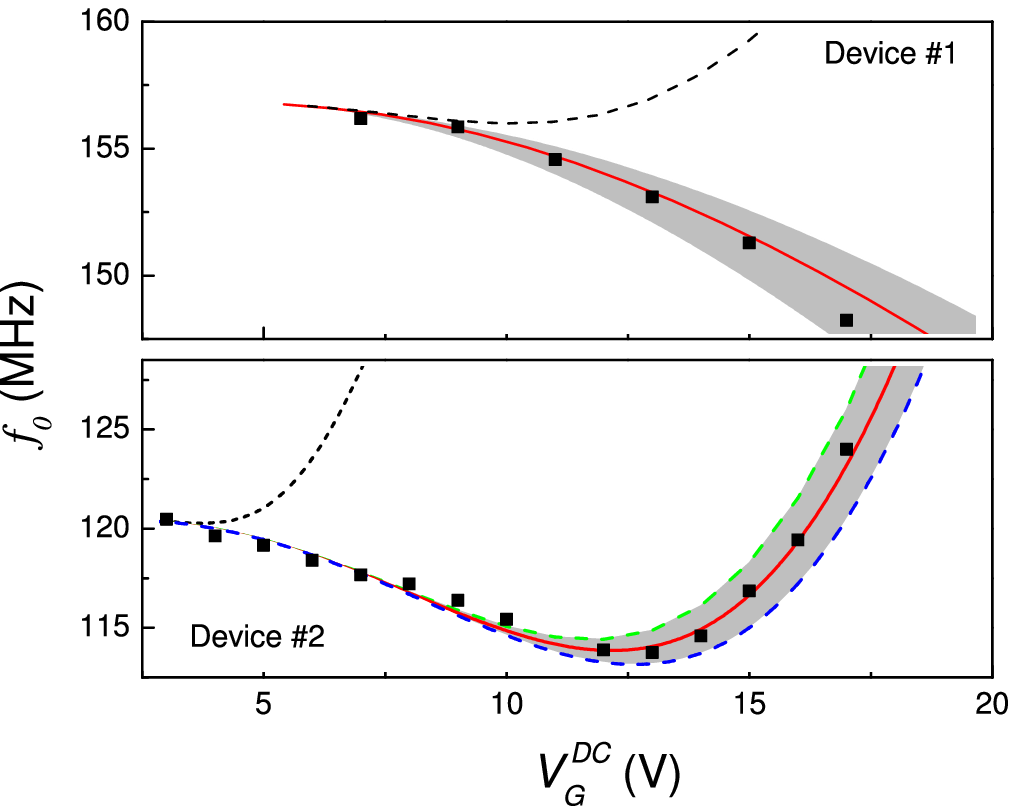}

\caption{\label{fig4}Resonant frequency as a function of $V{}_{G}^{DC}$ extracted
from the best fit of $I^{LI}\left(\omega\right)$ for device \#1 (shown
in \ref{fig3}) and device \#2, $L=800~\mathrm{nm}$ and $1000$~nm
respectively. Black dash curves: simulation using $T_{0}=0$. Red
curves: best fits $E=\left(3\pm1.5\right)\mathrm{GPa}$ $T_{0}=\left(2\pm0.1\right)\mu N$
for device \#1 $E=\left(10\pm5\right)\mathrm{GPa}$ $T_{0}=\left(1\pm0.1\right)\mu N$
for device \#2. Gray regions represent the impact of thickness $H$
uncertainty for device \#1 and the sensitivity to the value of $E$
for device \#2. Dash green curve: $E=11\mathrm{\: GPa}$ $T_{0}=1\mu{\rm N.}$
Dash blue curve: $E=9\mathrm{\: GPa}$ $T_{0}=1\mu{\rm N}$ .}

\end{figure}

Figure \ref{fig4} shows the extracted resonant frequency for two
different devices: \#1 and \#2 ($L=800\mathrm{nm}$ and $1000$nm
respectively). As expected, the resonant frequency extrapolated to
zero gate voltage is higher for device \#1 since the array length
is smaller. For device \#2 we observe a decrease at low gate voltages
and a strong increase at higher voltages.

To explain the observed dependence, we consider the SWNTs array as
an effective beam \cite{Comment2} and we use a continuum model to
describe its electromechanical behavior. Such approach was succefully
used in the case of individual MWNTs in Ref.{[}28{]}. If we assume
that the driving electrostatic force does not change the shape of
lowest vibration mode of the SWNTs array $u_{1}\left(x\right)$, the
vertical displacement of the SWNTs array $y\left(x,t\right)$ can
be approximated by $y_{0}\left(x\right)+u_{1}\left(x\right)\cdot\delta y\left(t\right)$,
i.e. a stationary part obeying the beam equation determined by the
DC electrostatic force and a small time dependent part $\delta y\left(t\right)$
obeying the equivalent harmonic oscillator equation (see Supplementary
Information (SI) for details)

\begin{equation}
\fl M\:\frac{d^{2}\delta y}{dt^{2}}+\mu\:\frac{d\delta y}{dt}+K\cdot\delta y=\tilde{C}_{G}^{'}V{}_{G}^{DC*}v_{G}\cdot\cos\left(\omega t\right)+\tilde{C}_{G}^{''}\left(V_{G}^{DC*}\right)^{2}\cdot\delta y\label{eq:harmonic}\end{equation}

in \eref{eq:harmonic} the spring constant $K$ is determined by
the physical and geometrical parameters of the SWNTs array and the
mechanical tension set by $V_{G}^{DC}.$ As can be expected, the spring
constant $K$ is an increasing function of $V_{G}^{DC}$ since the
mechanical tension increases with $V_{G}^{DC}.$ The value of $K$
can be obtained by solving self-consistently the beam equation (see
SI for details). The second term on the right-hand side of \eref{eq:harmonic}
comes from the non-linear dependence of the gate capacitance $C_{G}$
with the vertical coordinate. This term renormalizes the spring constant
and produces a softening of the resonator. Then the resonant frequency
of the SWNTs array can be written as follows (see SI for details)

\begin{eqnarray}
\fl f_{0}\left(V{}_{G}^{DC\,}\right)=\frac{1}{2\pi}\sqrt{\frac{K-\tilde{C}_{G}^{''}\left(V_{G}^{DC*}\right)^{2}}{M}}=\label{eq:freq vs Vg}\\
\fl\sqrt{\; f_{1}^{2}\left(1+\alpha\tilde{T_{0}}\right)-\beta\cdot\left(V_{G}^{DC*}\right)^{2}+f_{1}^{2}\alpha\intop_{0}^{1}\left(\frac{dy_{0}}{dx}\right)^{2}dx+f_{1}^{2}\kappa\left(\intop_{0}^{1}\frac{d^{2}y_{0}}{dx^{2}}u_{1}\left(x\right)dx\right)^{2}}\nonumber \end{eqnarray}

where $f_{1}\simeq1.03\frac{H}{L^{2}}\sqrt{E/\rho}$ corresponds to
the fundamental resonant frequency of a doubly clamped beam, $\alpha\simeq0.1475$,
$T_{0}=\tilde{T_{0}}\left(EWH^{3}/2L^{2}\right)$ is the mechanical
residual tension when $V_{G}^{DC*}=0$, $\beta=\varepsilon_{0}/(4\pi^{2}\rho H\cdot H_{0}^{3})$
and $\kappa\cong0.0166$. The third and forth terms in \eref{eq:freq
vs Vg} are the DC gate voltage dependent hardening of the effective
spring constant due to the elastic deformation of the beam. Numerical
simulations show that the sum of these two terms can be written as
$\lambda(E,\tilde{T_{0}})\cdot\left(V_{G}^{DC*}\right)^{4}$. As a
result, the proposed model fits quantitatively our data if $\beta\gg\lambda(E,\tilde{T_{0}})$.

AFM characterization of our devices shows that the thickness of the
SWNTs array is inhomogeneous. The mean value of the SWNTs membrane
thickness is about $\left(20\pm10\right)\mathrm{nm}$ for both devices.
The imperfect packing and alignment of SWNTs, that produces the mechanical
cohesiveness of the SWNTs membrane, makes its density unknown. However,
from data at low gate voltages one could estimate the value of $\beta$
and use its definition to obtain an effective density $\rho_{eff}$.
This yields: $\left(700\pm400\right)\mathrm{Kg\cdot m^{-3}}$ for
device \#1 and $\left(400\pm200\right)\mathrm{Kg\cdot m^{-3}}$ for
device \#2. Using this estimation and $E$ and $T_{0}$ as fitting
parameters one could obtain the results depicted in Figure \ref{fig4}.
For comparison we also show the best result considering $T_{0}=0$
(black dash curve) and using only $E$ as fitting parameter (120 GPa
and 610 GPa for device \#1 and \#2 respectively). The red curves correspond
to the best fit: $E=\left(3\pm1.5\right)\mathrm{GPa}$, $T_{0}=\left(2\pm0.1\right)\mu\mathrm{N}$
for device \#1 and $E=\left(10\pm5\right)\mathrm{GPa}$, $T_{0}=\left(1\pm0.1\right)\mu\mathrm{N}$
for device \#2. Gray region represents the uncertainty on the thickness
$H$ for device \#1 and the sensitivity to the value of $E$ for device
\#2. The quantitative agreement makes evident the finite value of
an in-built mechanical tension. This mechanical tension can probably
be attributed to the suspension step of SWNTs membranes. The impact
of dielectrophoresis field on $T_{0}$ will be further investigated.
The value obtained for the Young's modulus is better than the one
reported for bucky paper \cite{coleman2003} 2.3 GPa and comparable
to the one reported for highly aligned SWNTs \cite{Hayamizu2008}
or slightly smaller \cite{Hayamizu2009}, supporting that our SWNTs
arrays of thickness as small as 15nm behaves as highly ordered wafers.

The quality factor obtained from fitting $I^{LI}\left(\omega\right)$
is $35-55$ for both devices. This value is comparable to the one
reported by Hayamizu \emph{et al.} \cite{Hayamizu2009} for highly
aligned SWNTs thick cantilevers ($H\cong250\mbox{nm}$) and suspended
graphene \cite{Chen2009} but lower than the one reported for single
arc discharged SWNTs \cite{Dai1996}. One possible explanation is
related to the energy loss due to loosely linked SWNTs inside the
membrane. This value could be improved by suppressing the sliding
among SWNTs via cross-linking between them, as suggested in {[}32{]}.

In conclusion, SWNT suspended membrane gates were succesfuly integrated
to MOS transistors and operated at high frequency for the first time.
These extremely thin membranes behave as a cohesive, rigid and elastic
body. In particular, the resonant frequency of SWNTs SG-SiFETs is
tuned by the DC gate voltage and its dependence can be well described
within the continuum mechanics framework. The mechanical properties
of SWNTs arrays are comparable to those observed in closely packed
and aligned SWNTs wafers \cite{Hayamizu2008}. The quality factor
observed makes SWNTs membranes more suitable for high frequency mechanical
switching devices because of the quick damping of vibration. This
work shows promise to realize devices with practical application using
CNT, for example in bio-sensing and RF electronics.

\ack
This work was supported by Nano-RF, FP6 European project. We thank P. Joyez and C. Urbina for crittical reading of the manuscript.

\clearpage{}

\clearpage{}

\clearpage{}

\begin{thebibliography}{32}
\bibitem{CommentVGint} SI describes the procedure used to estimate
its value.
\bibitem{Comment}The phase difference between the gate and source
electrode depends on the frequency of the rf signals, on the (gate
dependent) conductance of the nanotube, and on its derivative. Moreover,
the lock-in detected phase does not contain any information about
$\varphi$.
\bibitem{Comment2}As a first approximation we consider the membrane
as a beam neglecting the correction due to the Poisson's ratio. This
is justified by the small value generally observed for the shear modulus
in this material.
\end{thebibliography}
\end{document}